\documentstyle[12pt]{article}

\begin{document}
\title{ Analytical representations based on $su(3)$ coherent states and Robertson
intelligent states }
\author{{\bf M. Daoud}{\footnote{Permanent adress: LPMC, Faculty of Sciences, University Ibn Zohr, Agadir,
 Morocco}}   \\
\\
Max Planck Institute for the Physics of Complex Systems\\ Dresden, Germany\
\\}

\maketitle

\begin{abstract}
Robertson intelligent states which minimize the Schr\" odinger-Robertson uncertainty
relation are constructed as eigenstates of a linear combination of Weyl generators
of the $su(3)$ algebra. The construction is based on the analytic representations of
$su(3)$ coherent states. New classes of coherent and squeezed states are explicitly
derived.

\end{abstract}

\vfill
\newpage 

\section{Introduction}
 The coherent states introduced as displacement of the ground state of harmonic oscillator by Schr\" odinger [1],
 have found revived
 interest when it was realized that they are eigenfunctions of the annihilation operator and minimize the
 Heisenberg uncertainty relation. The generalization of the usual coherent states from the Weyl-Heisenberg
 to the
 other Lie algebras and from harmonic oscillator to other potentials, followed these three approaches, namely,
 (i) eigenstates of lowering group generator for Lie algebras or annihilation operator of exactly solvable
  system,
  (ii) as orbits of the extremal weight state or (iii) as states minimizing the uncertainty relation. These different
 approaches lead to
 distinct sets of coherent states and coincide only in the special case of the harmonic oscillator
 (see [2-4] for review).
 Concerning  the optimization of the uncertainty principle, it was observed that a relation more accruate
 than Heisenberg one may be used to construct generalized coherent states and squeezed states. Indeed
 this relation known as Schr\" odinger-Robertson uncertainty inequality
 [5] can be minimized and gives rise to new sets of coherent and squeezed states ( see the pioneering works [6-8]). The
 states resulting from this minimization have different names in the literature such as correlated states [6-8] or
 Robertson intelligent states [9].\\
 More recently, there has been much interest in such states
 for Lie algebras [9-13] as well as for quantum systems evolving in various potentials [14-17]. Robertson intelligent states
 for the quadrature components of Weyl generators of the algebras $su(1,1)$ and $su(2)$ were constructed [9-13]. They were
 also defined for exactly solvable quantum systems as the eigenstates of complex combination of creation and
 annihilation operators [14-17].\\
 The purpose of this paper is to further extend the classes of Robertson intelligent states
 for higher symmetries.
 In this sense, the main idea of this note is to construct the intelligent states for
 the quadrature components of Weyl operators of the algebra $su(3)$. For this end, it may be useful
 to start by giving the
 explicit computation of the associated coherent states and their analytic representations. Hence, one can
 introduce
  the differential realizations for the $su(3)$ generators. As we will see, the analytic realization enables us  to convert the eigenvalue equations
 arising from the minimization of Schr\" odinger-Robertson inequality into quasi linear differential equations
   which provide
  the Robertson intelligent states.\\
  The paper is organized as follows. In section 2 we review the derivation of $su(3)$ coherent states.
   We compute
  explicitly the action of unitary displacement operator on the highset weight vector of the finite dimension representation
  space of the algebra $su(3)$. We give  the analytic representations of the $su(3)$ coherent states.
  We construct also the differential operators corresponding to the actions of the generators of $su(3)$ on
   the Fock-Bargmann space. In section 3, we show how the analytic representation based on the coherent states
  provides us
  with the tool to solve the eigenvalue equations resulting from the minimization of Schr\" odinger-Robertson
   relation and
  to obtain intelligent states for the quadrature components of $su(3)$ Weyl generators. We conclude in section 4
  after pointed out a number of interesting open problems.
\section{Analytic representations of $su(3)$-coherent states}
\subsection{Reconstructing $su(3)$-coherent states}
We shall begin by reexamining the construction
of the coherent states associated with a quantum system of dynamical symmetry $su(3)$.
 Althought this subjet has been considered previously in the works [18-19], we thought that is always interesting
 to give another method based
on the explicit computation of the action of displacement operator on the highset weight state (fudicial vector)
of the finite dimensional representation space of the algebra $su(3)$. The explicit forms of such states are needed
to perform their analytic representations and to give the realization of the generators of the algebra under
consideration. \\
The algebra $su(3)$ is defined by the generators $e_i$, $f_i$, $h_i$ ($i = 1, 2$) and the relations
\begin{equation}
[ e_i , f_i ] = \delta_{ij} h_j
\end{equation}
\begin{equation}
[ h_i , e_j ] = a_{ij} e_j {\hskip 0.5cm} [ h_i , f_j ] = -a_{ij} f_j
\end{equation}
\begin{equation}
[ e_i , e_j ] = 0 \qquad \textrm{for} \qquad \vert i - j \vert > 1
\end{equation}
\begin{equation}
e_i^2e_{i \pm 1} - 2 e_ie_{i \pm 1}e_i + e_{i \pm 1}e_i^2 = 0
\end{equation}
\begin{equation}
f_i^2f_{i \pm 1} - 2 f_if_{i \pm 1}f_i + f_{i \pm 1}f_i^2 = 0
\end{equation}
where $(a_{ij})_{i,j=1,2}$ is the Cartan matrix of $su(3)$, i.e. $a_{ii} = 2$, $a_{i,i\pm 1} = -1$
and $a_{ij} = 0$ for
$\vert i-j \vert > 1$. Many aspects of Lie algebras are best considered after choosing a special type
 of the representation basis. Since one would write down the $su(3)$ coherent states, the most convenient choice
 , in this case, is the bosonic realization. Indeed, an adapted basis is given in term of
  three bosonic pairs of creation and annihilation operators; they satisfy the commutation relations
\begin{equation}
[ a_k^{-} , a_l^{+}] = \delta_{kl}
\end{equation}
where $k,l = 1, 2, 3$. The operators number are $N_k = a_k^+a_k^-$. The Fock space is generated by the eigenstates
$\vert n_1, n_2, n_3\rangle$ of number operators, namely,
\begin{equation}
\vert n_1, n_2, n_3\rangle = \frac{(a_1^+)^{n_1}}{\sqrt{n_1!}}\frac{(a_2^+)^{n_2}}{\sqrt{n_2!}}\frac{(a_3^+)^{n_3}}{\sqrt{n_3!}}
\vert 0, 0, 0\rangle
\end{equation}
In this bosonic representation, we define the generators of $su(3)$ as
\begin{equation}
e_i = a_i^+ a_{i+1}^- {\hskip 0.5cm} f_i = a_i^- a_{i+1}^+ {\hskip 0.5cm} h_i = N_i - N_{i+1}
\end{equation}
The generators $e_i$, $f_i$ are called step, ladder or Weyl operators. The Cartan subalgebra is generated by
the elements $h_i$.
They act on the representation space of dimension $\frac{1}{2}(j_1+1)(j_1+2)$ that is obtained from the Fock space of three harmonic oscillators
by restricting the total number of quantas to $j_1 = n_1 + n_2 + n_3$. In the present representation the state
of highest weight is $\vert j_1, 0, 0\rangle$. The generators of $su(3)$ having a nontrivial action
( non-vanishing and non-diagonal) on the fudicial vector
$\vert j_1, 0, 0\rangle$ are $f_1 = a_1^-a_2^+$ and $f_3 = [f_2 , f_1] = a_1^-a_3^+$. At this stage, one can define
the coherent state as
\begin{equation}
\vert z_1 , z_2 \rangle = D(z_1 , z_2) \vert j_1, 0, 0\rangle
=\exp ( z_1 f_1 + z_2 f_3 - \bar z_1 e_1 - \bar z_2 e_3) \vert j_1, 0, 0\rangle
\end{equation}
where $e_3 = [e_1 , e_2] = a_1^+a_3^-$. Expanding the displacement operator $D(z_1 , z_2)$ and using the action of creation and annihilation
operators on the restricted Fock space \\
${\cal F} = \{\vert n_1, n_2, n_3 \rangle ; n_1+n_2+n_3 = j_1\}$, one get
\begin{equation}
\vert z_1 , z_2 \rangle = \sum_{j_2=0}^{j_1} \sum_{j_3=0}^{j_2}
z_1^{j_2} z_2^{j_3} I_{j_2}^{j_1}(|z_1|)I_{j_3}^{j_2}(|z_2|)
\vert j_1-j_2, j_2-j_3, j_3\rangle.
\end{equation}
where
\begin{equation}
I_{j_{s+1}}^{j_{s}}(|z_s|) = \sum_{k=0}^{\infty}\frac{(-)^k(|z_s|^2)^k}{(j_{s+1}+2k)!}
P(j_{s+1}+1,k),
\end{equation}
for $s = 1, 2$. The quantities $P$ occuring in (11) are given by
\begin{equation}
P(j_{s+1}+1,k) = P(j_{s+1}+1,0)\sum_{l_{1}=1}^{j_{s+1}+1}
E_s(l_1)\sum_{l_{2}=1}^{l_1+1}
E_s(l_2)...\sum_{l_{k}=1}^{l_{k-1}+1}
E_s(l_k)
\end{equation}
with $P(j_{s+1}+1,0) = \frac{j_s!j_{s+1}!}{(j_s - j_{s-1})!}$ and $E_s(l) = (j_s - l + 1)l$.
 They satisfy the following recursion relation
\begin{equation}
P(j_{s+1}+1,k) = \sqrt{E_s(j_{s+1})} P(j_{s+1},k) + \sqrt{E_s(j_{s+1}+1)} P(j_{s+1}+2,k-1).
\end{equation}
Setting
\begin{equation}
J^{j_s}_{j_{s+1}}(|z_s|) = |z_s|^{j_s}P(j_{s+1}+1,0) I^{j_s}_{j_{s+1}}(|z_s|),
\end{equation}
we get the first order differential equation
\begin{equation}
\frac{dJ^{j_s}_{j_{s+1}}(|z_s|)}{d|z_s|} = J^{j_s}_{j_{s+1}-1}(|z_s|) - (E_s(j_{s+1}+1))^2 J^{j_s}_{j_{s+1}+1}(|z_s|).
\end{equation}
The solution of this equation takes the simple form
\begin{equation}
J^{j_s}_{j_{s+1}}(|z_s|) = \frac{1}{j_{s+1}!} (\ cos (|z_s|))^{j_{s+1}-1}(\ tg (|z_s|))^{j_{s+1}},
\end{equation}
and the $su(3)$ coherent states rewrite as
\begin{eqnarray}
\lefteqn{\vert \zeta_1 , \zeta_2\rangle = (1 + |\zeta_1|^2 + |\zeta_1|^2|\zeta_2|^2)^{-\frac{j_1}{2}}} \nonumber\\
& & {}\times \sum_{j_2=0}^{j_1} \sqrt{\frac{j_1!}{j_2! (j_1 - j_2)!}}
\zeta_1^{j_2}  \sum_{j_3=0}^{j_2} \sqrt{\frac{j_2!}{j_3! (j_2 - j_3)!}}
\zeta_2^{j_3}
\vert j_1 - j_2, j_2 - j_3, j_3\rangle
\end{eqnarray}
where $\zeta_s = \frac{z_s}{|z_s|} \ tg (|z_s|) \cos (|z_{s+1}|)^{2-s}$ for $s= 1, 2$. They have
the property of strong continuity in the label space and overcompletion in the sense
that there exists a positive measure such that they solve the resolution to identity. The appropriate form of this resolution is
\begin{equation}
\int d\mu (\zeta_1 , \bar \zeta_1, \zeta_2 , \bar \zeta_2) \vert \zeta_1, \zeta_2\rangle \langle \zeta_1 , \zeta_2\vert =
\sum_{j_2=0}^{j_1} \sum_{j_3=0}^{j_2}\vert j_1 - j_2, j_2 - j_3, j_3\rangle \langle j_1 - j_2, j_2 - j_3, j_3 \vert.
\end{equation}
Assuming the isotropy of the measure $d\mu (\zeta , \bar \zeta)$,  we set
\begin{equation}
d\mu (\zeta_1 , \bar \zeta_1, \zeta_2 , \bar \zeta_2) = \pi^2 (1 + |\zeta_1|^2 + |\zeta_1|^2|\zeta_2|^2)^{\frac{j_1}{2}}
 h( \vert \zeta_1 \vert ^2) h( \vert \zeta_2 \vert ^2)
 \vert \zeta_1\vert d \vert \zeta_1\vert  \vert \zeta_2\vert d \vert \zeta_2
 \vert d\theta_1 d\theta_2
\end{equation}
with $\zeta_s = \vert \zeta_s \vert e^{i\theta_s}$.
Substituting (19) in Eq.(18), we obtain the following sum
\begin{equation}
\int_{0}^{\infty} x_s^{j_{s+1}} h(x_s) dx_s = \frac{j_{s+1}!(j_{s}-j_{s+1})!}{j_s!}.
\end{equation}
which should be satisfied
by the function $h(x_s = \vert \zeta_s \vert ^2 ))$.
One get
\begin{equation}
h(x_s) = \frac{j_{s}+1}{(1+x_s^2)^{j_s+2}}.
\end{equation}
This result can be obtained by using the definition of Meijer's $G$-function and
the Mellin inversion theorem [20]. The resolution to identity is necessary to build up the Fock-Bargamann
space based on the set of $su(3)$ coherent states.
\subsection{ Differential realization of the $su(3)$ generators}
It is well established that  the use of the Fock-Bargmann representation is a powerful method for
obtaining closed analytic expressions for various properties of coherent states. Calculation for some
quantum exceptation values and solutions for some eigenvalue equations are simlpified by exploiting the theory
of analytical entire functions. Here, we give the Fock-Bargamnn representation of a $su(3)$ quantum mechanical
system. We define the Fock-bargamnn space as a space of functions which are holomorphic. The scalar product
is written with an integral of the form
\begin{equation}
\langle f \vert g \rangle = \int {\bar f(\zeta _1, \zeta _2)}
g(\zeta _1, \zeta _2) d\mu (\zeta_1 , \bar \zeta_1, \zeta_2 , \bar \zeta_2)
\end{equation}
where the measure is defined above (see Eq.(19)). Due to overcompletion of the coherent sates,
it is induced by the scalar
product in ${\cal F}$. Let
\begin{equation}
\vert \psi \rangle = \sum _{n_1,n_2,n_3} a_{n_1,n_2,n_3}\vert n_1,n_2,n_3\rangle
\end{equation}
an arbitrary quantum state of ${\cal F}$, it can be represented
as a function of the complex variables $\zeta_1, \zeta_2$ as
\begin{equation}
\psi (\zeta_1, \zeta_2) = (1 + |\zeta_1|^2 + |\zeta_1|^2|\zeta_2|^2)^{\frac{j_1}{2}}
\langle \bar \zeta_1, \bar \zeta_2 \vert \psi \rangle
\end{equation}
 In particular, the analytic functions associated to elements of the basis of ${\cal F}$ are defined as
\begin{equation}
\psi _{j_1, j_2, j_3}(\zeta_1, \zeta_2) = (1 + |\zeta_1|^2 + |\zeta_1|^2|\zeta_2|^2)^{\frac{j_1}{2}}\langle \bar \zeta_1, \bar \zeta_2
\vert j_1 - j_2, j_2 - j_3, j_3 \rangle.
\end{equation}

We now investigate the form of the action of the operators $e_i$, $f_i$ and $h_i$ on Fock-Bargmann space.
Indeed, any operator $O$ of the algebra $su(3)$ is represented in the space of entire analytical functions by some differential operator
${\cal O}$, defined by
\begin{equation}
\langle \bar \zeta_1, \bar \zeta_2 \vert O \vert \psi \rangle = {\cal O} \psi (\zeta_1, \zeta_2)
\end{equation}
for any state $\vert \psi \rangle $ of ${\cal F}$.\\
According this definition, we obtain
\begin{equation}
e_1 = \frac{\partial}{\partial \zeta_1} {\hskip 0.5cm} e_3 = \frac{\partial}{\partial \zeta_2}
\end{equation}
\begin{equation}
f_1 = j_1 \zeta_1 - \zeta_1^2\frac{\partial}{\partial \zeta_1} - \zeta_1\zeta_2\frac{\partial}{\partial \zeta_2}
\end{equation}
\begin{equation}
f_3 = j_1 \zeta_2 - \zeta_2^2\frac{\partial}{\partial \zeta_2} - \zeta_1\zeta_2\frac{\partial}{\partial \zeta_1}
\end{equation}
\begin{equation}
e_2 = \zeta_1\frac{\partial}{\partial \zeta_2} {\hskip 0.5cm} f_2 = \zeta_2\frac{\partial}{\partial \zeta_1}
\end{equation}
\begin{equation}
h_1 = j_1 - 2 \zeta_1\frac{\partial}{\partial \zeta_1} - \zeta_2\frac{\partial}{\partial \zeta_2}
\end{equation}
\begin{equation}
h_2 = \zeta_1\frac{\partial}{\partial \zeta_1} - \zeta_2\frac{\partial}{\partial \zeta_2}.
\end{equation}
To obtain the above differential realization:\\
\indent (i) we remark that the coherent states (17) can be also written as
\begin{equation}
\vert \zeta _1, \zeta _2 \rangle = (1 + |\zeta_1|^2 + |\zeta_1|^2|\zeta_2|^2)^{\frac{-j_1}{2}}
D(\zeta _1, \zeta _2) \vert j_1 , 0, 0 \rangle
\end{equation}
where $D(\zeta _1, \zeta _2) = \exp (\zeta _1f_1 + \zeta _2f_3)$,\\
\indent (ii) we observe that
\begin{equation}
\frac{\partial}{\partial \zeta _1}D(\zeta _1, \zeta _2) = f_{1}D(\zeta _1, \zeta _2)
{\hskip 0.5cm}
\frac{\partial}{\partial \zeta _2}D(\zeta _1, \zeta _2) = f_{3}D(\zeta _1, \zeta _2),
\end{equation}

\indent (iii) we use the Hausdorff formula
\begin{equation}
e^{-B} A e^B = \sum _{n \geq 0} \frac{1}{n!} \big( - adB\big)^n A
\end{equation}
where $ (adB)A = [ B, A]$, \\
\indent (iv) we use also the actions of the elements of $su(3)$ on the basis of Fock space ${\cal F}$, in particular the
fudicial vector $\vert j_1, 0, 0 \rangle$, and the structure relations (1-5) of the algebra $su(3)$.\\
From the previous considerations, it follows that the $su(3)$ generators  act as
 first-order holomorphic differential operators on the space of the analytic functions generated by the
 elements (25). One can verify that the
commutation relations (1-5) are preserved. This result combined with eigenvalue equations ensuring the minimization
of Schr\" odinger-Robertson inequality provides the intelligent states as that will be explained in the next section.

\section{\bf $su(3)$ Robertson states}
As we have already mentionned, in this section, we will study
 the fluctuations of the quadrature components of Weyl generators which represent creation and annihilation
 of states for a quantum mechanical system of $su(3)$ symmetry. In this order, to construct the intelligent
 states of any pair of ladder operators $e_i$, $f_i$ ($i=1, 2, 3$), it is natural to introduce
 the quantum observables  $ \sqrt{2}p_i = e_i + f_i$ and
${\bf i}\sqrt{2}q_i = e_i - f_i$  where ${\bf i}^2=-1$. These observables obey
\begin{equation}
[p_i , q_i ] = {\bf i} h_i.
\end{equation}
 We known that $p_i$ and $q_i$ satisfy, in a given state, the Robertson-Shr\" odinger
uncertainty
relation
\begin{equation}
(\Delta p_i)^2 (\Delta q_i)^2 \geq \frac{1}{4} ( \langle h_i\rangle^2 + \langle c_i\rangle^2),
\end{equation}
where $\Delta p_i$ and $\Delta q_i$ are the dispersions and the hermitian operator$ c_i = \{ p_i - \langle p_i\rangle, q_i - \langle q_i\rangle\}$
gives the covariance (correlation) of the observables $p_i$ and $q_i$. The symbol $\{ , \}$ stands for the standard definition of the
anticommutator. A state $\vert \Phi \rangle$ providing the equality in (37) is
the so-called Robertson intelligent state. It was proven that such state satisfy the following
eigenvalue equation
\begin{equation}
\big((1 + \alpha) e_i + (1 - \alpha) f_i \big) \vert \Phi \rangle =  \lambda \vert \Phi \rangle
\end{equation}
where $\alpha \neq 0$ and $\lambda = (1 + \alpha) \langle e_i\rangle + (1 - \alpha) \langle f_i\rangle$
are complex parameters. Furthermore, the variances and covariance , in the intelligent state
$\vert \Phi \rangle$, are related by
\begin{equation}
(\Delta p_i)^2 = \vert \alpha \vert \Delta _i  {\hskip 0.5cm}  (\Delta q_i)^2 = \frac{1}{\vert \alpha \vert} \Delta _i
\end{equation}
where $ \Delta _i = \frac{1}{2} \sqrt{ \langle h_i\rangle^2 + \langle c_i\rangle^2}$.
Remark that they can be also expressed as
\begin{equation}
(\Delta p_i)^2 = \frac{\vert \alpha \vert^2}{u}\langle h_i\rangle {\hskip 0.5cm}
(\Delta q_i)^2 = \frac{1}{u}\langle h_i\rangle {\hskip 0.5cm}
\langle c_i\rangle = \frac{v}{u} \langle h_i\rangle
\end{equation}
where the real parameters $u$ and $v$ are such that $u^2 + v^2 = 4 \vert \alpha \vert^2$ ( As example, one can take $ u = 2Re\alpha$ and $v = 2Im\alpha$
). It is clear that the dispersions
and the correlation can be obtained from the mean value of the observable $h_i$.
The state $\vert \Phi \rangle $ satisfying
 (38) with $\vert \alpha \vert = 1$ are coherent because they satisfy $(\Delta p_i)^2
  = (\Delta q_i)^2 = \Delta _i$. The fluctuations are equals and minimized
  in the sense of Schr\" odinger-Robertson uncertainty relation. The state satisfying
  (38) with $\vert \alpha \vert \neq 1$ are squeezed because if $\vert \alpha \vert  < 1$,
  we have $(\Delta p_i)^2 < \Delta _i < (\Delta q_i)^2$ and if  $\vert \alpha \vert > 1$,
  we have $(\Delta q_i)^2 < \Delta _i < (\Delta p_i)^2$.
  \\
\indent To solve the eigenvalues equation (38), we will use the anlytic representations of coherent states as well as
the differential realizations of the the generators $e_i$ and \\$f_i$ given by Eqs.(27-30).
 So, let us start by deriving the eigenfunctions of Eq.(38) for the first pair  $e_1$, $f_1$.
 By introducing the analytic function
 \begin{equation}
 \Phi _1 \equiv \Phi _1( \zeta _1,\zeta _2, \alpha , \lambda , j_1) = (1 + |\zeta_1|^2 + |\zeta_1|^2|\zeta_2|^2)^{\frac{j_1}{2}}
\langle \bar \zeta_1, \bar \zeta_2 \vert \Phi _1 \rangle,
 \end{equation}
 it can be easly checked that the eigenvalue equation (38) can be converted in the following first order
  differential equation
\begin{equation}
(j_1 \eta_1 - \lambda ') \Phi _1 + ( 1 - \eta _1^2 )\frac{\partial \Phi _1}{\partial \eta_1} - \eta _1 \eta_2 \frac{\partial \Phi _1}{\partial
\eta_2}  = 0,
\end{equation}
where $ \eta _1 = \sqrt{\frac{1-\alpha}{1+\alpha}} \zeta_1$, $\eta _2 = \zeta_2$ and $\lambda'= \frac{\lambda}{\sqrt{1-\alpha ^2}}$
for $ \alpha \neq \pm 1 $. The function
$ \Phi _1( \zeta _1,\zeta _2, \alpha , \lambda , j_1)$  can be expanded as
\begin{equation}
\Phi _1 = \sum _{j_2=0}^{j_1}\sum_{j_3=0}^{j_2} a_{j_1,j_2,j_3} \eta _1^{j_2} \eta _2^{j_3}.
\end{equation}
Substitution of (43) in (42) yields the recursion formula
\begin{equation}
(j_1 - j_2 - j_3 + 1) a_{j_1,j_2 - 1,j_3} - \lambda ' a_{j_1,j_2,j_3} + (j_2 + 1) a_{j_1,j_2 + 1,j_3} = 0
\end{equation}
which  can be solved
by the Laplace method. Indeed, we set
\begin{equation}
a_{j_1,j_2,j_3} = \int_{-1}^{+1} x^{j_2} f(x) dx
\end{equation}
that we introduce in (44) to obtain, after partial integration, the simple first order differential equation
satisfied by the function $f(x)$
\begin{equation}
(x - x^3) \frac{df}{dx} +  (j_1 - j_3 + 1 - \lambda 'x - x^2) = 0.
\end{equation}
The last equation is easly solvable. Replacing in (45), one get
\begin{equation}
a_{j_1,j_2,j_3} = \int _{-1}^{+1} x^{j_2-j_1+j_3-1} (1-x)^{\frac{-\lambda'+j_1-j_3}{2}}(1+x)^{\frac{\lambda'+j_1-j_3}{2}}dx,
\end{equation}
or
\begin{eqnarray}
\lefteqn{a_{j_1,j_2,j_3} = (-)^{j_2} \frac{\Gamma (\frac{\lambda'+j_1-j_3}{2} + 1)\Gamma (\frac{-\lambda'+j_1-j_3}{2} + 1)}
{\Gamma (j_1 - j_3 + 2)}} \nonumber\\
& &{} \times {_2}F_1(j_1 - j_3 - j_2 + 1, \frac{\lambda'+j_1-j_3}{2} + 1, j_1 - j_3 + 2, 2)
\end{eqnarray}
using the integral representation for the hypergeometric function $_2F_1$ [20]. Comparing the expansion (43)
 with the general
formula (41), we have the decomposition of Robertson intelligent states over the basis of Fock space ${\cal F}$
\begin{equation}
\vert \Phi _1 \rangle = \sum _{j_2=0}^{j_1}\sum_{j_3=0}^{j_2} a_{j_1,j_2,j_3} \Big(\frac{1-\alpha}{1+\alpha}\Big)^{\frac{j_2}{2}}
\sqrt{\frac{(j_1-j_2)! j_3!(j_2-j_3)!}{j_1!}} \vert j_1-j_2, j_2-j_3, j_3\rangle
\end{equation}
where the coefficients $a_{j_1,j_2,j_3}$ are given by Eq.(48).\\
\indent Now we consider the construction of intelligent states for the second pair $e_2$,$f_2$.
The eigenvalues equation (38)
 gives, in this case, the following quasi linear differential equation
\begin{equation}
\xi_1 \frac{\partial \Phi_2}{\partial \xi_2} + \xi_2 \frac{\partial \Phi_2}{\partial \xi_1} - \lambda'\Phi_2 = 0
\end{equation}
where $\xi_1 = \sqrt{\frac{1+\alpha}{1-\alpha}}\zeta_1$, $\xi_2 = \zeta_2$ and $\lambda' = \frac{\lambda}{\sqrt{1-\alpha^2}}$.\\
Here also, we expand the eigenfunction $\Phi _2 \equiv \Phi _2( \xi_1,\xi_2, \alpha , \lambda , j_1)$ as
\begin{equation}
\Phi_2 = \sum_{j_2=0}^{j_1}\sum_{j_3=0}^{j_2} b_{j_1,j_2,j_3}\xi_1^{j_2}\xi_2^{j_3}
\end{equation}
that we insert in the equation (50) to obtain the recursion relation linking the coefficients $b$'s
\begin{equation}
(j_3 + 1)b_{j_1,j_2-1,j_3+1} - \lambda'b_{j_1,j_2,j_3} + (j_2 + 1) b_{j_1,j_2+1,j_3-1} = 0.
\end{equation}
Setting $ b_{j_1,j_2,j_3} \equiv b_{j_1, j_2 - j, j}$ where $ 2j = j_2+j_3$, the previous relation can be transformed to
\begin{equation}
(j_2 + 1) b_{j_1, j_2 - j + 1, j} - \lambda'b_{j_1, j_2 - j, j} + (j_3  + 1)b_{j_1, j_2 - j - 1, j} = 0,
\end{equation}
sovable in a similar manner that one given the solution of recursion formula (44), and one has
\begin{eqnarray}
\lefteqn{b_{j_1,j_2,j_3} = (-)^{j_2} \frac{\Gamma (\frac{\lambda'+j_2+j_3}{2} + 1)\Gamma (\frac{-\lambda'+j_2+j_3}{2} + 1)}
{\Gamma (j_2 + j_3 + 2)}} \nonumber\\
 & & {}\times{_2}F_1(j_3 + 1, \frac{\lambda'+ j_2 + j_3}{2} + 1, j_2 + j_3 + 2, 2)
\end{eqnarray}
Finally, one obtain
\begin{equation}
\vert \Phi _2 \rangle = \sum _{j_2=0}^{j_1}\sum_{j_3=0}^{j_2} b_{j_1,j_2,j_3} \Big(\frac{1+\alpha}{1-\alpha}\Big)^{\frac{j_2}{2}}
\sqrt{\frac{(j_1-j_2)! j_3!(j_2-j_3)!}{j_1!}} \vert j_1-j_2, j_2-j_3, j_3\rangle
\end{equation}
\indent It remains to determine the $e_3$,$f_3$ intelligent states. In this case, the Robertson states should satisfy the
following equation
\begin{equation}
(j_1 \vartheta_2 - \lambda ') \Phi _3 + ( 1 - \vartheta _2^2 )
\frac{\partial \Phi _3}{\partial \vartheta_2} - \vartheta _1 \vartheta_2 \frac{\partial \Phi _3}{\partial
\vartheta_1}  = 0
\end{equation}
where $\vartheta _1 = \zeta_1$, $ \vartheta _2 = \sqrt{\frac{1-\alpha}{1+\alpha}}\zeta_2$ and $\lambda'$ is defined above.
In a similar way that one presented above, one obtain the intelligent states
\begin{equation}
\Phi_3 = \sum_{j_2=0}^{j_1}\sum_{j_3=0}^{j_2} c_{j_1,j_2,j_3} \vartheta_1^{j_2}\vartheta_2^{j_3}
\end{equation}
where $\Phi _3 \equiv \Phi _3( \zeta _1,\zeta _2, \alpha , \lambda , j_1)$ and the constants $c$'s are given by
\begin{eqnarray}
\lefteqn{c_{j_1,j_2,j_3} = (-)^{j_3} \frac{\Gamma (\frac{\lambda'+j_1-j_2}{2} + 1)\Gamma (\frac{-\lambda'+j_1-j_2}{2} + 1)}
{\Gamma (j_1 - j_2 + 2)}}\nonumber\\
 & & {}\times {_2}F_1(j_1 - j_2 - j_3 + 1, \frac{\lambda'+j_1-j_2}{2} + 1, j_1 - j_2 + 2, 2).
\end{eqnarray}
Analogously to the above cases, the intelligent states $\Phi _3$ can be converted as follows
\begin{equation}
\vert \Phi _3 \rangle = \sum _{j_2=0}^{j_1}\sum_{j_3=0}^{j_2} c_{j_1,j_2,j_3} \Big(\frac{1+\alpha}{1-\alpha}\Big)^{\frac{j_3}{2}}
\sqrt{\frac{(j_1-j_2)! j_3!(j_2-j_3)!}{j_1!}} \vert j_1-j_2, j_2-j_3, j_3\rangle.
\end{equation}
To close this section, let us note that it is clear that the Fock-Bargmann representation of
the coherent states
provide a simplification and a "minimization" in the problem of finding intelligent
states of $su(3)$ Weyl generators.
It is evident that the procedure described here can be relevant in the derivation of intelligent states for
other quadrature components of type, for instance, $e_i$, $e_j$ and $f_i$, $f_j$ ($i \neq j$).

\section{\bf Discussion and outlook}
 In conclusion, we have developed a method for finding the Robertson intelligent
 states for linear combination of the Weyl operators $e_i$ and $f_i$ for $i=1, 2, 3$,
 corresponding to the Lie algebra $su(3)$. The use of the analytic representation
 enables us to write the eigenvalue equations, satisfied by states minimizing the  Schr\" odinger-Robertson
 uncertainty relation, as quasi linear first order differential equation. Interestingly,
 new types of coherent states for $su(3)$ emerge for $\vert \alpha \vert = 1$.
 Also when $\vert \alpha \vert \neq 1$, the solutions give squeezed states.
 As it is noted in the end of the previous section, the approach used through this work can be applied to derive Robertson intelligent
 states associated to the other quadratures of the $su(3)$ generators. In an unified scheme, they can be obtained
 by considering the eigenvalue problem for an operator which is a complex linear combination of all elements
  of $su(3)$
 \begin{equation}
 \sum_{i=1,2} \big(\alpha_i^+ e_i + \alpha_i^- f_i + \alpha_i^0 h_i \big) \vert \Phi \rangle = \lambda \vert \Phi \rangle
  \end{equation}
 The solutions of such general problem give the so-called in the literature algebra
 eigenstates or algebraic coherent states ([13] and references therein). Taking specific constraints on the complex parameters
 occurring in this general eigenvalue equation, one can get various kind of coherent and
 squeezed states, in particular ones not discussed in this paper. This constitutes
 the first possible prolonogation of our results. Also, as continuation, it would be interesting
 to apply the approach presented here to other Lie algebras like $su(n)$ or
 $su(p,q)$.
 \vfill\eject

{\bf Acknowledgements}:
The author would like to acknowledge the Max Planck Institute for the Physics of Complex
Systems for kind hospitality and helpful atomosphere .\\

{

\end{document}